\documentclass[letterpaper, 10 pt, conference]{ieeeconf}  % Comment this line out if you need a4paper

\IEEEoverridecommandlockouts                              % This command is only needed if you want to use the \thanks command

\usepackage{graphicx}
\usepackage{textcomp}
\usepackage{xcolor}
\usepackage{hyperref}
\usepackage{subcaption}
\title{\LARGE \bf
Adaptive Learning based Upper-Limb Rehabilitation Training System with Collaborative Robot
}

\author{Jun Hong Lim, Kaibo He, Zeji Yi, Chen Hou, Chen Zhang, Yanan Sui, Luming Li % <-this % stops a space
\thanks{National Engineering Research Center of Neuromodulation, Tsinghua University, Beijing, China {\tt\small \{linjh19,hkb21,yizj20,hou-c17\} @mails.tsinghua.edu.cn,\{zhangchen2020,ysui,lilm\} @tsinghua.edu.cn}}%
}

\begin{document}

\nocite{*}
\maketitle
\thispagestyle{empty}
\pagestyle{empty}

%%%%%%%%%%%%%%%%%%%%%%%%%%%%%%%%%%%%%%%%%%%%%%%%%%%%%%%%%%%%%%%%%%%%
\begin{abstract}

Rehabilitation training for patients with motor disabilities usually requires specialized devices in rehabilitation centers. Home-based multi-purpose training would significantly increase treatment accessibility and reduce medical costs. While it is unlikely to equip a set of rehabilitation robots at home, we investigate the feasibility to use the general-purpose collaborative robot for rehabilitation therapies. In this work, we developed a new system for multi-purpose upper-limb rehabilitation training using a generic robot arm with human motor feedback and preference. We integrated surface electromyography, force/torque sensors, RGB-D cameras, and robot controllers with the Robot Operating System to enable sensing, communication, and control of the system. Imitation learning methods were adopted to imitate expert-provided training trajectories which could adapt to subject capabilities to facilitate in-home training. Our rehabilitation system is able to perform gross motor function and fine motor skill training with a gripper-based end-effector. We simulated system control in Gazebo and training effects (muscle activation level) in OpenSim and evaluated its real performance with human subjects. For all the subjects enrolled, our system achieved better training outcomes compared to specialist-assisted rehabilitation under the same conditions. Our work demonstrates the potential of utilizing collaborative robots for in-home motor rehabilitation training.

\textit{Clinical relevance—}The collaborative robot system is capable of providing safe and effective training comparable to specialized rehabilitation robots, enabling possibilities of convenient rehabilitation training at home.
\end{abstract}

%%%%%%%%%%%%%%%%%%%%%%%%%%%%%%%%%%%%%%%%%%%%%%%%%%%%%%%%%%%%%%%%%%%%%%%%%%%%%%%%
\section{INTRODUCTION}

Rehabilitation training is essential for the recovery of patients with motor disabilities. Patients need to go through intensive and repetitive training conducted by an experienced therapist to build up their muscle strength and neuroplasticity in dedicated rehabilitation centers using specialized devices. %For neurological diseases and injuries such as stroke, survivors suffer from a series of sequelae, including movement and language disorders. Rehabilitation 
It is the primary way for them to regain motor functions and return to normal life. For patients suffering from neurological diseases and injuries such as stroke, upper-limb motor impairments are common sequelae. Training should begin immediately once the patient is medically stable and rehabilitation goals can be established.%are able to be identified for rehabilitation. 
It can help strength weak muscles to prevent muscle atrophy and recover motor functions, avoiding hand edema or shoulder joint subluxation due to lack of movement. 

%In the acute period of stroke, the patient is weak in muscle strength and requires passive assistance, it also assists in preventing muscle atrophy and maintaining muscle functions so that it does not lead to hand edema or shoulder joint subluxation due to lack of movement. Motor recovery could be accelerated with proper training, and various outpatient clinical trials adopt a training of an hour per day, three times a week for a period of two to three months.

% Improve Figure 1 and Figure 2.

\begin{figure}[thpb]
      \centering
       \includegraphics[scale=0.35]{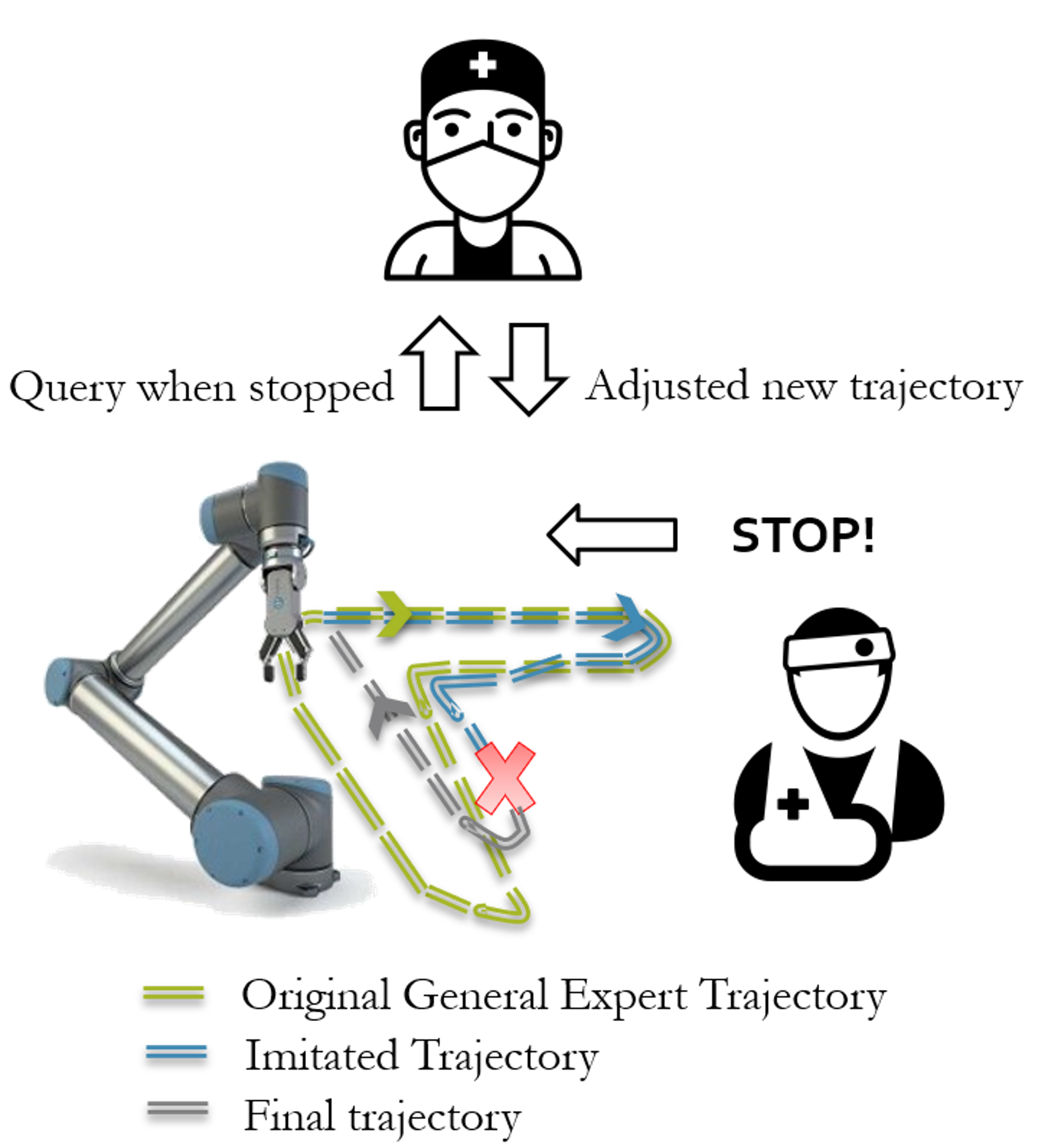}
       \caption{Training adapted to subject capabilities}
       \label{fig:idea}
   \end{figure}

%\section{Related Work}
Robotic systems, including both dedicated and general-purpose robots, are widely used for rehabilitation as they are capable of performing repetitive tasks consistently. %We review dedicated and general-purpose rehabilitation robots for the upper limb. 
Dedicated robots are usually designed to train for a certain task, while general-purpose robots can be utilized in various scenarios, have a greater degree of freedom compared to dedicated systems and are more cost-effective.
%equipped with less degree of freedom, and are less cost-effective. Collaborative robots can be utilized in various scenarios, not just rehabilitation training, therefore they can be more cost-effective and possess a larger degree of freedom as compared to dedicated devices. 

%\subsection{Rehabilitation Robotics for Upper-Limb}

%Rehabilitation training improves capabilities such as self-care, mobility, communications, cognitive, and social skills. It should start immediately when the patient is medically stable and able to benefit from the training. Training details vary at different stages, it usually involves a therapist assisting the patient with designed exercises. Robotic systems are frequently used in rehabilitation. They play an important part by helping to make the training more independent, efficient, and interesting.

%Over the past half-century, 
Several types of dedicated robotic rehabilitation devices and control systems for upper-limb rehabilitation had been developed over the past decades \cite{chen2016assistive,aprile2020upper,mazzoleni2017combining}.
%reviewed typical assistive control systems for upper-limb rehabilitation robots. 
These rehabilitation robots and assistive control systems can guide patients through the training process with or without therapist surveillance. They can also help collect training data for the therapist to evaluate the current conditions of the patient.
\begin{figure*}[htbp]
    \centering
       \includegraphics[scale=0.43]{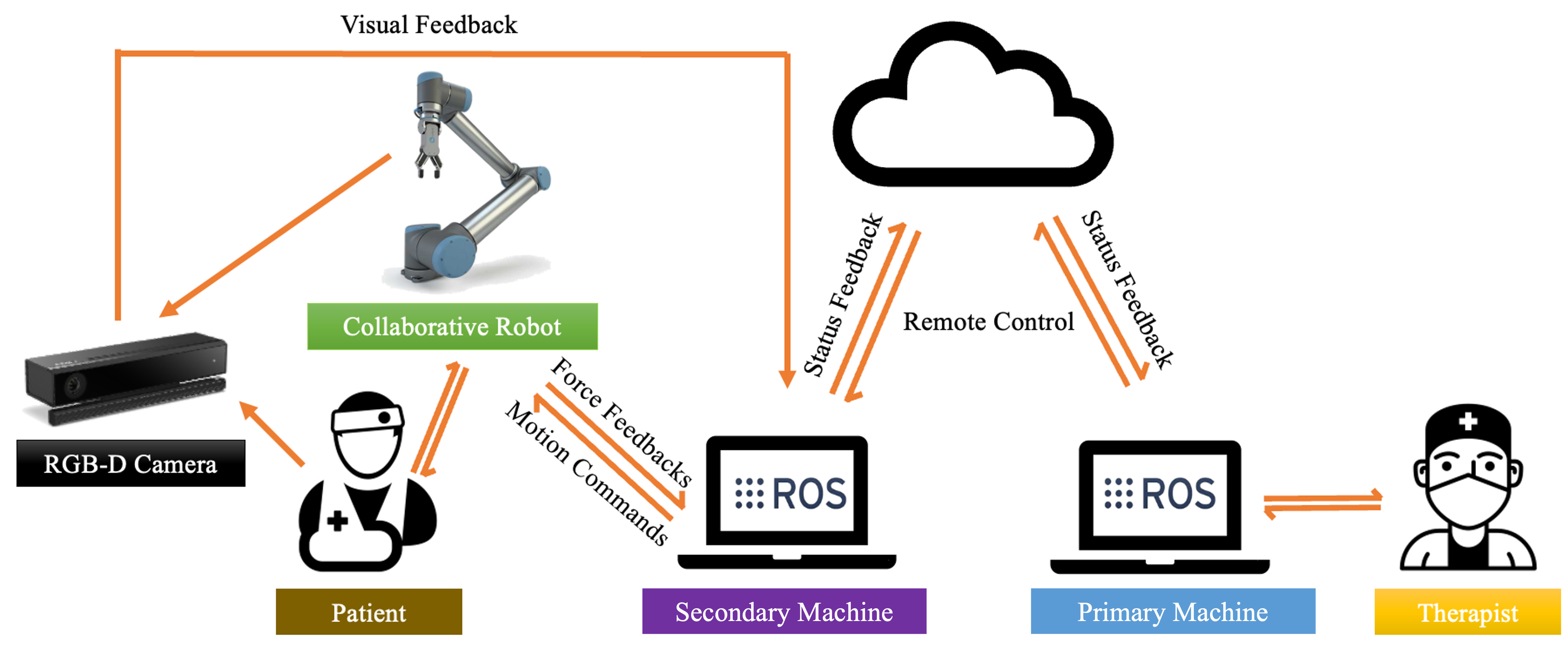}
       \caption{Overview of the proposed remote system}
       \label{fig:system}
   \end{figure*}

\label{subsec:cobotRehab}
There are limited attempts on investigating the application of general-purpose collaborative robots for rehabilitation tasks. Collaborative robots (Cobots) can work together with humans and often comply with safety standards, showing the potential to be deployed in rehabilitation processes. Kyrkjeb{\o} et al. \cite{kyrkjebo2018feasibility} attempted to increase the safety of the Cobot by restraining workspaces in supplement to built-in safety mechanisms. Their results indicated the feasibility of using the collaborative robot to assist in rehabilitation training. Nielsen el al. integrated dynamic motion primitives with force feedback to pre-record and learn a task for rehabilitation, adapting to different arm lengths \cite{nielsen2017individualised}. Another work from \cite{becker2018controlling} used a KUKA robot arm with 7-DOF. The arm incorporated the assist-as-needed principle to motivate patients with different levels of assistance. All of the approaches suggested an primary way to enable repetitive tasks and that patients were fairly independent to practice daily activities with less supervision needed.

The time- and expense-consuming process of gaining access to rehabilitation centers and the shortage of manpower in areas with limited resources may delay the recovery phase of patients. %If remote training can be provided with similar training effects, these problems could be solved. 
%There is still limited research on investigating the potential usage of the general-purpose collaborative robot (Cobot) on assisting rehabilitation training. 
A possible way to solve this problem is to use a remote training system which provides comparable training effects. In this work, we propose a remote adaptive upper-limb training system with a collaborative robot. The system integrates Cobot, camera, movement sensors, therapist, and subject in one platform. It is able to perform different types of training and achieve similar training effectiveness as common rehabilitation robots, demonstrating the feasibility of deploying a general-purpose Cobot to facilitate personalized rehabilitation training at home instead of frequent visits to rehabilitation centers. %This telemedicine training can be further divided into gross level and fine level movement training, both are essential for rehabilitation.

\section{Methods}

We proposed a new remote system for assisted rehabilitation training and integrated with the ROS platform to enable communications among a set of hardware (shown in Figure~\ref{fig:system}). Gripper-based end-effector was utilized for guiding finer movement tasks that could not be performed with previous end-effector approaches.
%(Section~\ref{subsec:cobotRehab}). 
Subjects could be too weak to hold the gripper, so we developed this rehabilitation task with assistance from doctors and therapists, which is very important for rehabilitation training in early stroke rehabilitation. The proposed system was validated through simulations in Gazebo and OpenSim. We examined the exercise effects by surface electromyography (sEMG) signals with comparison to experienced trainers through experiments to reveal its feasibility in multi-purpose rehabilitation.

%Connecting and allowing communications between the therapist and the subject, remotely via ROS through the cloud. 
The subject were guided by Cobot's end-effector to complete the training. The movement, gestures, and rehabilitation duration were designed based on the stroke guidelines to care for acute phase patients\cite{hebert2016canadian,powers2018} with instructions, amendments, and recommendations from experienced physicians and therapists. Two tasks were proposed in this paper. One for gross level arm rehabilitation and another for fine level hand rehabilitation. For arm rehabilitation, we selected the basic ADL task, which is essential in daily life. For hand rehabilitation, we select the Finger training task in order to facilitate finger nimbleness. During the opening and closing of the fist, it actuated the entire hand and forearm muscle. Therefore, it provided better training efficiency for acute-stage patients, helping them to build up muscle strength. A piece of assisting equipment (as shown in Figure~\ref{fig:training}(e) and (f)) were worn on the subject's forearm, whose grip-site would be identified by computer vision methods\cite{bradski2000opencv} in real-time with a Kinect camera.
% The camera could also act as a surveillance to subjects' other physical conditions.

\begin{figure}[t!]
     \centering
     \begin{subfigure}[t]{0.1\textwidth}
         \centering
         \includegraphics[height=1.2in]{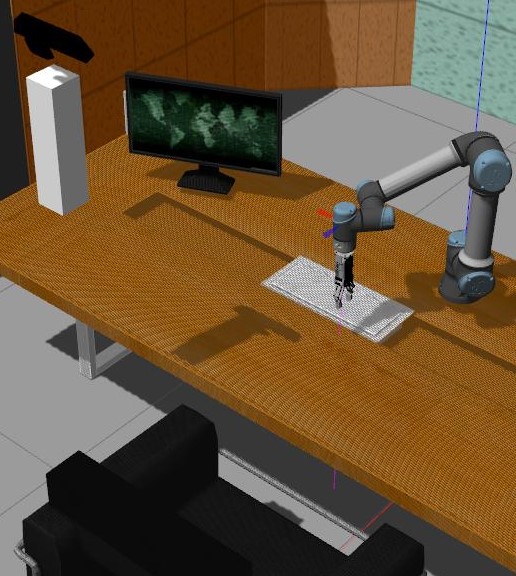}
         \caption{}
         %\label{fig:y equals x}
     \end{subfigure}
     \hfill
     \begin{subfigure}[t]{0.1\textwidth}
         \centering
         \includegraphics[height=1.2in]{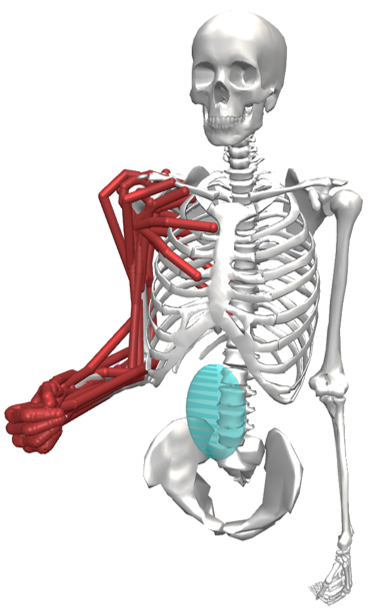}
         \caption{}
         %\label{fig:three sin x}
     \end{subfigure}
     \hfill
     \begin{subfigure}[t]{0.1\textwidth}
         \centering
         \includegraphics[height=1.2in]{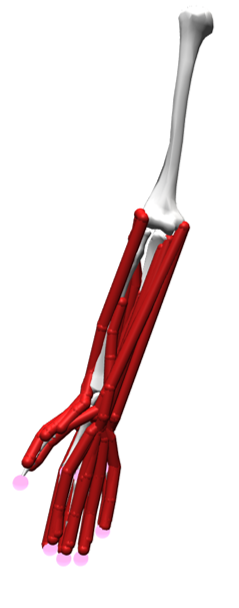}
         \caption{}
         %\label{fig:five over x}
     \end{subfigure}
        \caption{(a) Simulation of the training system with visual and motor feedback in Gazebo. (b) Simulation of gross level motor training with OpenSim. (c) Simulation of fine-level motor training with OpenSim.}
        \label{fig:opensim}
\end{figure}

%\subsection{Robotic Simulation Model and Safety Configurations}
\label{simulationthreshold}
%Our collaborative robot arm platform is an UR5e from Universal Robot. A Force/Torque (F/T) sensor, gripper, and a Kinect camera were also integrated into the simulation environment as shown in Figure~\ref{fig:opensim}(a). 

A Gazebo simulation environment was built as shown in Figure~\ref{fig:opensim}(a). The camera was setup to have a clear view of the workspace, RGB-D images were transmitted to our system for recognizing and the grip-site of the subject's assist equipment were identified. The resisting force due to body limitations such as the range of motion were tracked by the Force/Torque sensor and reported simultaneously to the controller for analysis. We fixed a resistant force threshold for emergency, if $F\geq F_{safe}$ (45N), the Cobot triggers an emergency stop as further action may lead to additional injuries. We adopted the starting points for each subject based on visual recognition. The simulation was utilized to verify task feasibility, the imitation learning agent also use the environment to learn an adaptive control policy using general expert data. Expert and imitated trajectories were simulated in Gazebo and OpenSim simulators to ensure their safety and evaluate the corresponding muscle activation rates in order to optimize the task as described in Section \ref{subsec:osim}.

%As the task is rich in human interaction, our main concern is to ensure its safety.
Safety is our main concern for rich human-interactive tasks. On top of factory default safety configurations, the Cobot safety was configured to three stages, determined by the subject's physical abilities, range of motion (ROM) of the rehab arm, and their stages of rehabilitation. They were different in workspace determination, acceleration, and velocity applied during the training. 
%We call Stage 1 the Super Critical Stage and is the safest. At this stage, subjects have the most limited ROM. Stage 2 is Critical, where the limits are gently eased to meet greater capabilities. Stage 3 is Safe, with the workspace limited to ROM that a normal person can reach and 40 \% of default speed. The subject can move on to harder stages as recovery progress.
%they can gradually move on to level three when training and recovery progress, where a wider ROM and a faster velocity and acceleration limits are provided.

%\subsection{Musculoskeletal Simulation}
\label{subsec:osim}
OpenSim\cite{delp2007opensim} is an open-sourced model-based biomechanical simulation software. It is widely used for human kinematics and dynamics analysis. We built two musculoskeletal models in OpenSim to verify that the target muscles can be activated effectively. As shown in Figure~\ref{fig:opensim}, for Arm Rehabilitation, we built an upper body model based on \cite{raabe2016investigation}. It mainly included the skeletons and muscles of hands, arms, and shoulders. For Hand Rehabilitation, we built an upper limb model, which included detailed and precise skeletons and muscles of the hand and forearm. 

In the simulation, we obtained the initial trajectory of the Cobot's end-effector (the initial gripper trajectory) from Gazebo simulations. Then we used this trajectory to drive the musculoskeletal model and muscle activation were calculated. According to the muscle activation results, We discussed with clinical experts to adjust the tasks, generates a new trajectory and the model were driven again to calculate muscle activation results. After several adjustments, we obtained the appropriate tasks and verify the effectiveness.

%\subsection{Task Design}

%Previous work in Section~\ref{subsec:cobotRehab} either use a customized end-effector or require the subject to hold on to it during the training, which both require a certain amount of muscle strength. Subjects could be too weak to hold the gripper, so we developed this rehabilitation task with assistance from doctors and therapists, which is very important for rehabilitation training in early stroke rehabilitation. The most important thing, as suggested, which is also stated in the guidelines\cite{hebert2016canadian,powers2018} is to strengthen their muscles and prevent muscle atrophy and maintain muscle function so that it does not lead to hand edema or shoulder joint subluxation due to lack of movement.

%Based on this principle, we designed an arm rehabilitation task simulating daily life's tasks and a more precise finger-movement training, the hand rehabilitation task. This is a highly recommended task by physicians. During the opening and closing of the fist, it actuates the entire hand and forearm muscle. Therefore, it provides better training efficiency for acute-stage patients, helping them to build up muscle strength.

%\subsection{Control Methods}
\label{subsec:control}
Imitation learning enables an agent to learn policies/trajectories from expert demonstrations. It has great potential to be applied in rehabilitation robotics. For remote training, therapists are not on-site to guide the Cobot on rehabilitation tasks. Instead, they can provide trajectories that are generally suitable for the subjects. This acts as an expert-labeled reference. The learning agent executes it on the subject side through imitation learning and adjusts with respect to physical conditions and personal preference.

%In autonomous vehicles researches, there are track boundaries to restrict the agent and warns them when they failed to keep themselves inside the boundaries. 
Unlike autonomous vehicles, there are no track boundaries to restrict the agent in rehabilitation training. Each subject has different boundaries determined by their biological conditions. In our approach, as shown in Figure \ref{fig:idea}, subjects can press ``stop" when the trajectory is too difficult for them. It is similar to violating track boundaries in autonomous driving. Based on this principle, we adopted the imitation methods inspired by HG-DAgger \cite{kelly2019}, with modifications to fit our problem.

%\begin{figure}[thpb]
      %\centering
      
       %\includegraphics[scale=0.25]{pic/figure4.png}
       %\caption{3D Trajectory simulated for one round of task in the experiment. (Red: Expert-provided trajectory Blue: Imitated Trajectory)}
       %\label{fig:traj}
   %\end{figure}
\begin{figure}[thpb]
      \centering
      
       \includegraphics[scale=0.34]{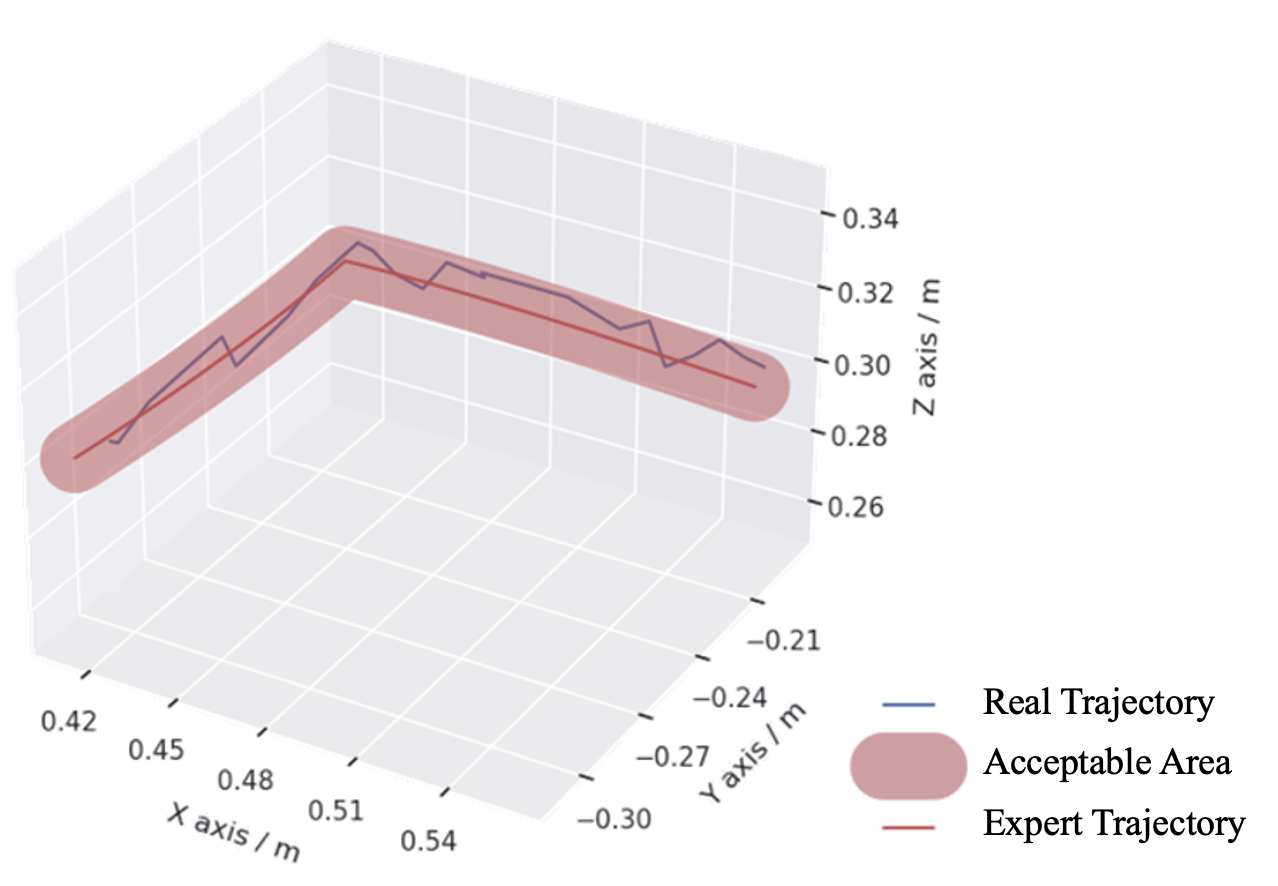}
       \caption{Real Trajectory(BLUE) executed for another new Expert Trajectory(RED LINE) from the trained model. The red area is the acceptable area by experts. RMSE is around $1e-2$m.}
       \label{fig:pred}
   \end{figure}

% In HG-DAgger, expert takes control when they feel necessary. 
In our system, the provided expert trajectory was deployed till a negative feedback had been flagged by the subject, and the training task was then adjusted by the experts. The stopping state would be marked as a bad state.
The states before were considered acceptable states. % We determine the stage before negative call as good state. 
Then, the expert will provide an adjusted consequent trajectory, and training resumes with the adjusted trajectory from that state. The procedure will be repeated until the subject accepts the entire trajectory. The final trajectory will be trained and export a policy personalized for the subject. 

%Like all other DAgger training, the novice policy is trained on a training dataset $D$, which is collected with expert labels during repeated rollouts either by simulating or by recording the trajectory while manipulating the Cobot. 
%Every state $S$ and Action $A$ in expert data set $D$ will be deployed. There are two conditions that a trajectory will be marked as a bad state $B$. First, if the trajectory is beyond the acceptable range for the subject, he/she can press the key to stop the Cobot. Second, the F/T sensor will continue to track resistance forces exerted by the subject, once it exceeds the threshold stated in section \ref{simulationthreshold}, the Cobot halts, and the state is recorded. The expert will adjust the original trajectory to reduce the difficulty, either through remote or by hand manipulation. Once the subject feels better, this indicates trajectory is now inside the \textit{SAFE REGION}. 
%The expert intervened data $D_j$ will be merged with data before bad state $D_B$. The control will then return to novice policy and resumes the training of the model. This model will be utilized to provide the subject guidance throughout the rest of the training sessions. 
%The trajectory for both training tasks (one round) is simulated as shown in Figure \ref{fig:traj}. 

The actual trajectory executed for a new expert data test set from the trained model were shown in Figure \ref{fig:pred}. The model was evaluated through RMSE with the distance respect to the trajectory ($x_g$,$y_g$,$z_g$) predicted and original expert trajectory ($x_i$,$y_i$,$z_i$).

\begin{figure}[b!]
    \centering
    \begin{subfigure}[t]{0.5\textwidth}
        \centering
        \includegraphics[width=\textwidth]{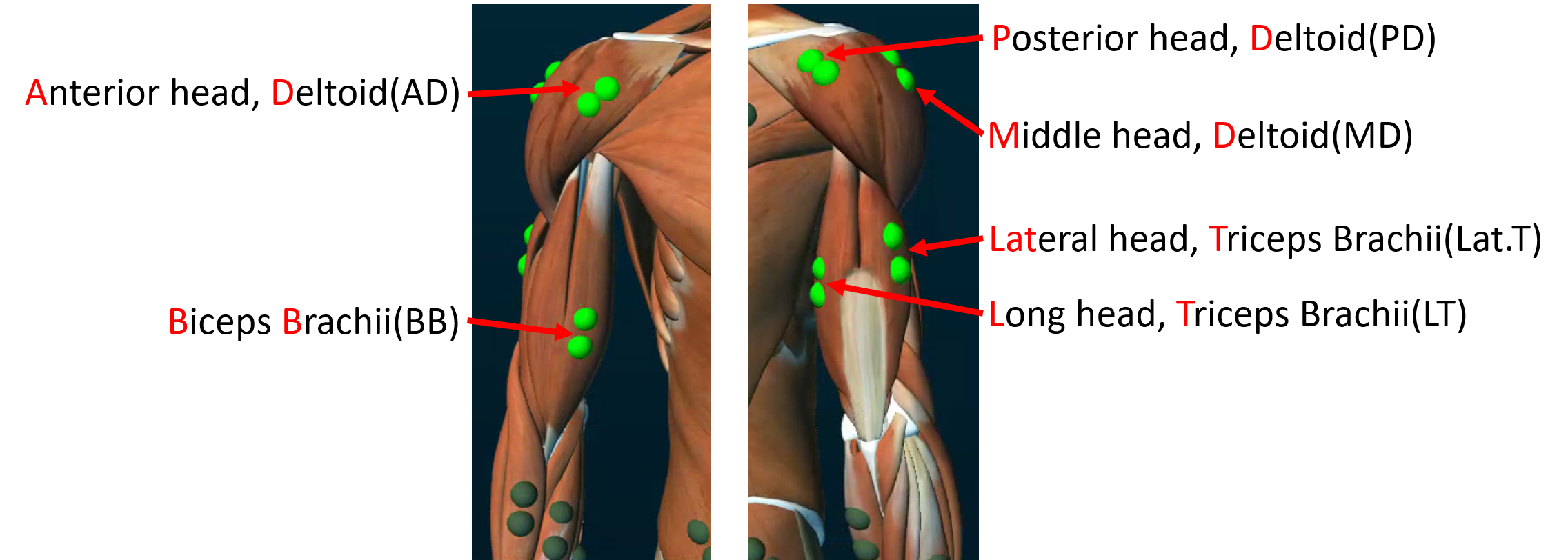}
        \caption{Arm Rehabilitation Muscle}
    \end{subfigure}
    \hspace{5pt}
    \begin{subfigure}[t]{0.5\textwidth}
        \centering
        \includegraphics[width=\textwidth]{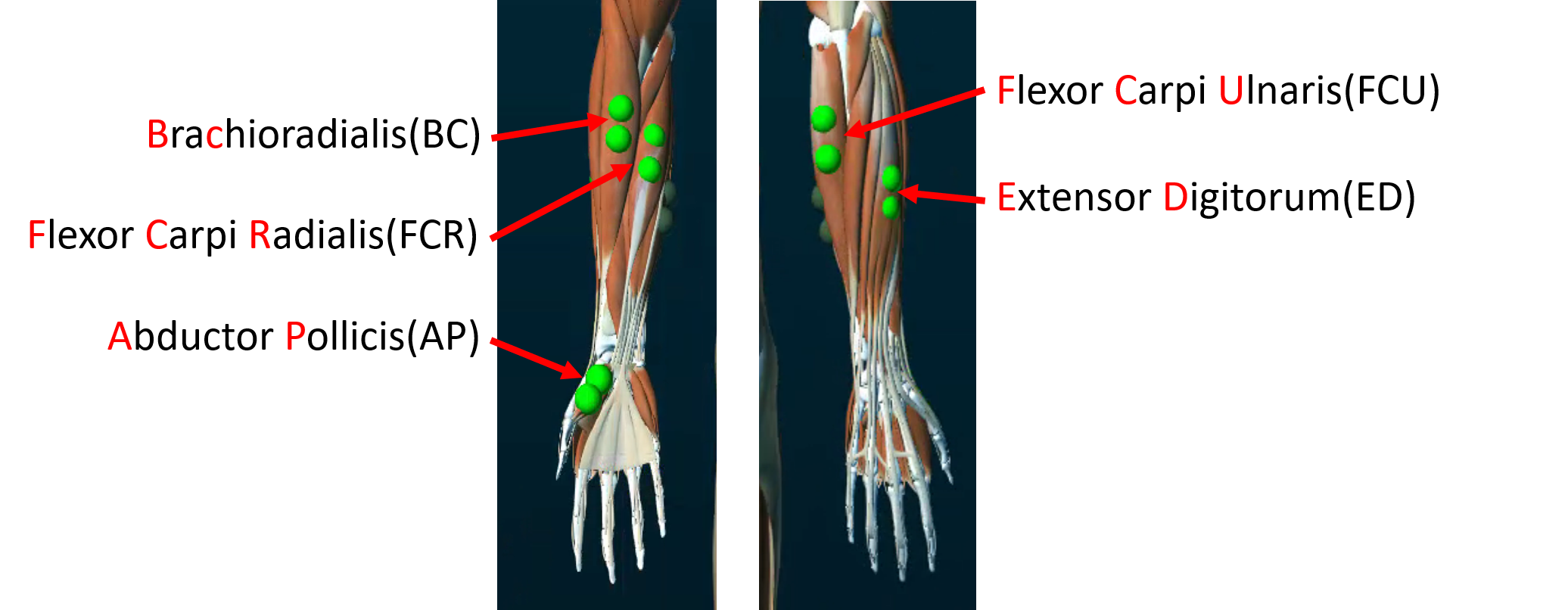}
        \caption{Hand Rehabilitation Muscle}
    \end{subfigure}
    \caption{Target muscles chosen to evaluate the training effects (Adopted from muscles illustrated in Noraxon MR3 program)}
    \label{fig:marker}
\end{figure}
%\subsection{Evaluation}

%To evaluate the effectiveness of rehabilitation training with the collaborative robot, we record surface electromyography (sEMG) data to get muscle activation rates from the subjects. 11 channels of sEMG signals are used to identify and analyze the training loads on different targeted muscle groups of the upper limb.

The feedback from the subject's perspective was also valuable as it directly reflects the feeling of interaction with the collaborative arm during training. To receive and study the feedback, we designed a survey to acquire their views after training with our rehabilitation system degree of exercises on arm and hand, safety measures taken, and a comprehensive rating for the overall task.

\section{Experiment and Results}

%\subsection{Experiments with Human Subjects}

%Due to pandemic restrictions, we were unable to carry out clinical experiments. Instead, 
We recruited five healthy subjects to evaluate the safety and feasibility of the training provided by our rehabilitation system. Initial expert trajectories for both tasks were determined by experts using simulators to emulate suitable trajectories, and verified using OpenSim and Gazebo simulators to ensure the safety execution on human subjects. All subjects were well informed and asked to be relaxed and remain passive during training. They were also informed to provide resistance force to imitate movement disorders so as to evaluate the self-adaptive system based on the F/T sensor and generate personalized training plans based on imitation learning.

To measure muscle activation quantitatively, We selected six target muscles associated with the movement of the upper limb for ADL task and five target muscles related to finger training task as shown in Figure~\ref{fig:marker} (a) and (b). 11 channels of sEMG data were measured and analyzed for each subject to validate if certain muscle activation rates had been achieved. Then the experiment was repeated with a trained specialist executing the same training task while holding on to the same attach point as the Cobot. The muscle activation rates are measured and used for comparison. 
%Then, a survey was conducted to reveal the subject's feelings regarding the human-robot interaction, safety, and opinions for the automated collaborative training task.

\begin{figure*}[htbp]
    \centering
       \includegraphics[scale=0.5]{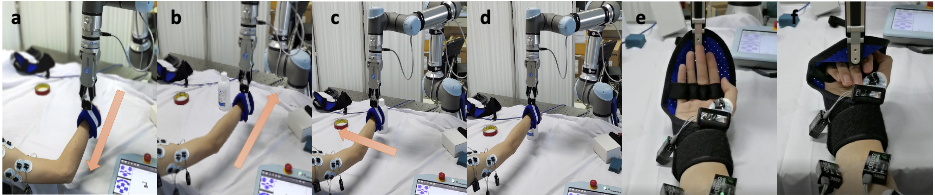}
       \caption{Subject receiving automated training with daily life task (Fig 8(a)(b)(c) Gross-level arm rehabilitation training) and precise finger training (Fig 8(d)(e) Fine level hand rehabilitation training) (a) Backward Contraction (b)Forward Extension (c) Arm Swivel Stretch (d) Pick Up with Guidance (e) Palm Open (f) Fist Hold}
       \label{fig:training}
   \end{figure*}

Two different cases of training were performed in the experiment, including the arm rehabilitation task and the hand rehabilitation task (as shown in Figure~\ref{fig:training}). Training was conducted at a 5-minute interval, as suggested by clinicians. A button was provided to stop the Cobot from executing when the subject thinks it is in \textit{BAD STATE}. If so, the expert will then take control as mentioned in Section \ref{subsec:control}. There was also a 20 seconds rest between each interval of the training. 
\begin{figure}[t!]
    \centering
    \begin{subfigure}[t]{0.45\textwidth}
        \centering
        \includegraphics[width=\textwidth]{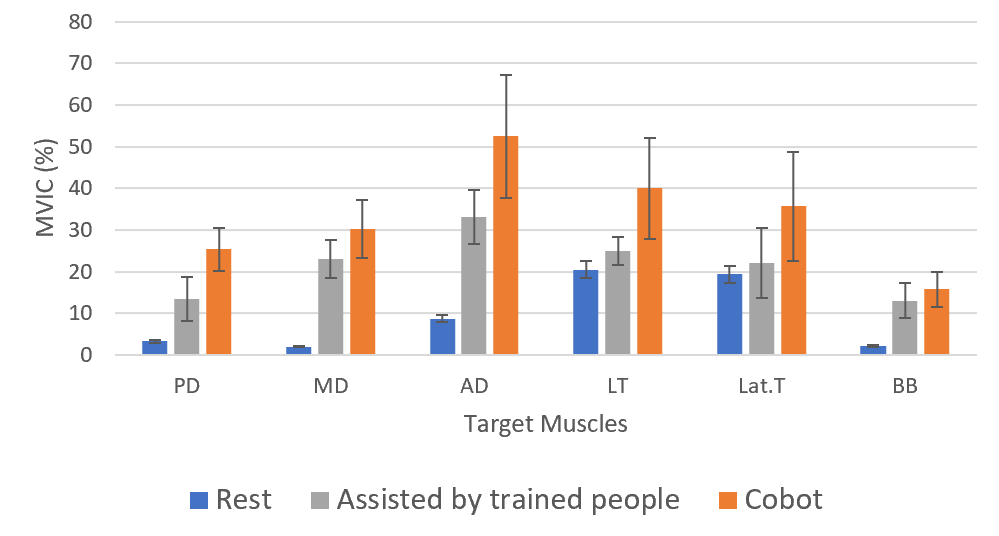}
        
        \caption{Muscle Activation in Gross-level training}
    \end{subfigure}
    
    \vspace{0.1 in}
    
    \begin{subfigure}[t]{0.45\textwidth}
        \centering
        \includegraphics[width=\textwidth]{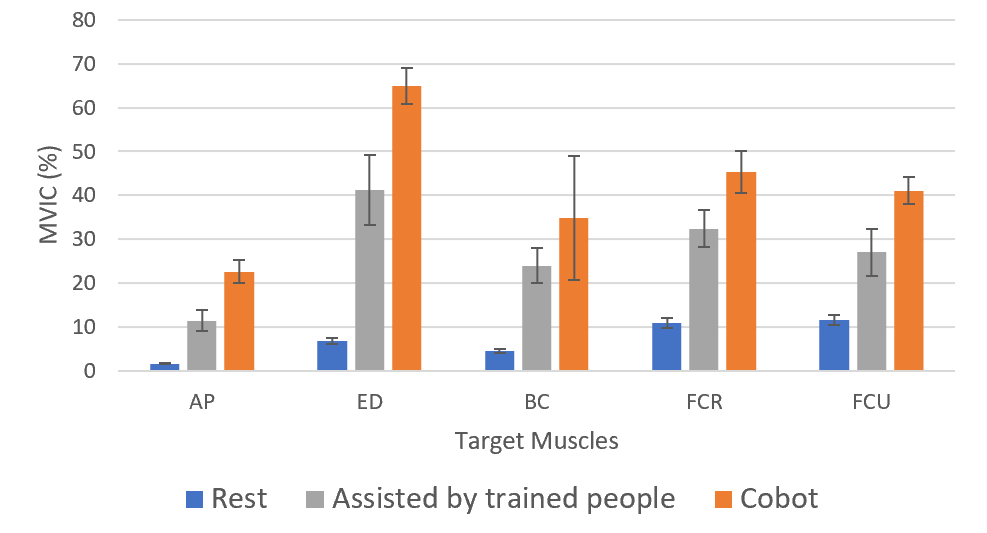}
        
        \caption{Muscle Activation in Fine-level training}
    \end{subfigure}
    
    \caption{Muscle activation based on raw-sEMG data for each subject when conducting rehabilitation training, shown in percentage of Maximum voluntary isometric contraction for each subject. (Muscle abbreviations as shown in Figure~\ref{fig:marker}.)}
    \label{fig:results}
\end{figure}

%\subsection{Results}
Wireless EMG sensors (Noraxon Inc.) were used and properly placed according to the guideline to measure muscle activation. The subject's sEMG at rest and during training was recorded and shown as a comparison in Figure~\ref{fig:results}. The data were smoothed (RMS at 100ms window), time normalized, averaged across repetitions, amplified, and ECG artifacts removed. 

We compared Cobot-guided muscle activation rates to specialist-assisted training (with the same attach point, trajectory, and instruction given to the subject). To normalize the sEMG values for comparison, we recorded the Maximum Voluntarily Isometric Contraction (MVIC) of each subject with identical posture and conditions and divide them by the measured sEMG values. MVIC is a common standardized method to measure the muscle strength of humans \cite{dara2007,lee2016,visser2003}.We averaged the values between all subjects as shown in Figure~\ref{fig:results}. The experiment result clearly shown that training on our system induced a significantly higher MVIC as compared to training with a specialist.

Compared with the muscle activation rate of specialists in assisted rehabilitation, each target muscle of Cobot assisted rehabilitation task shown a higher activation amount. Target muscles in gross-level training achieved an average of 11.6($ \pm$5.9)\% increment and fine-level training achieved an average of 14.6($\pm$5.3)\% increment.

The experiment results reflected a clear muscle activation during the designed training exercises, and the fist holding task is effective in activating the entire forearm muscles during training. In both tasks, the activate degrees of the subjects carried by the Cobot outweigh that by the therapist significantly. The following reasons may be essential:
\textbf{1.} The movement of the subject was more extended so that the Cobot could stretch the forearm fully in each training cycle. Thus, the performance was more stable.
\textbf{2.} The Cobot applied a larger force than the therapist. The therapist could retain force for a new subject as the physical properties are not known. While the Cobot was equipped with the adaptive method, it could actively adjust the strength given and allow subjects to their full limits.
\textbf{3.} The end-effector made the process more effective. The end-effector provided better support and containment. Hence, the force was more uniform, and a larger resultant force could be applied.

Besides the points listed above, there were other factors that can affect the training performance of therapists in real practice. In our setting, we gave the therapist plenty of rest between each task, and the training period was fixed to 10-min. However, a real therapist's workload is much higher. Therefore, fatigue is inevitable.
Alternating or even combining the tasks proposed will effectively help in training the entire arm muscles. Hence our proposed system is feasible and offered an efficient way to build muscle strength, and prevent contracture and deformation of the arm during the early stages of rehabilitation, where the subject might be weak to perform standard exercises voluntarily.

\begin{figure}[thpb]
      \centering
      
       \includegraphics[scale=0.14]{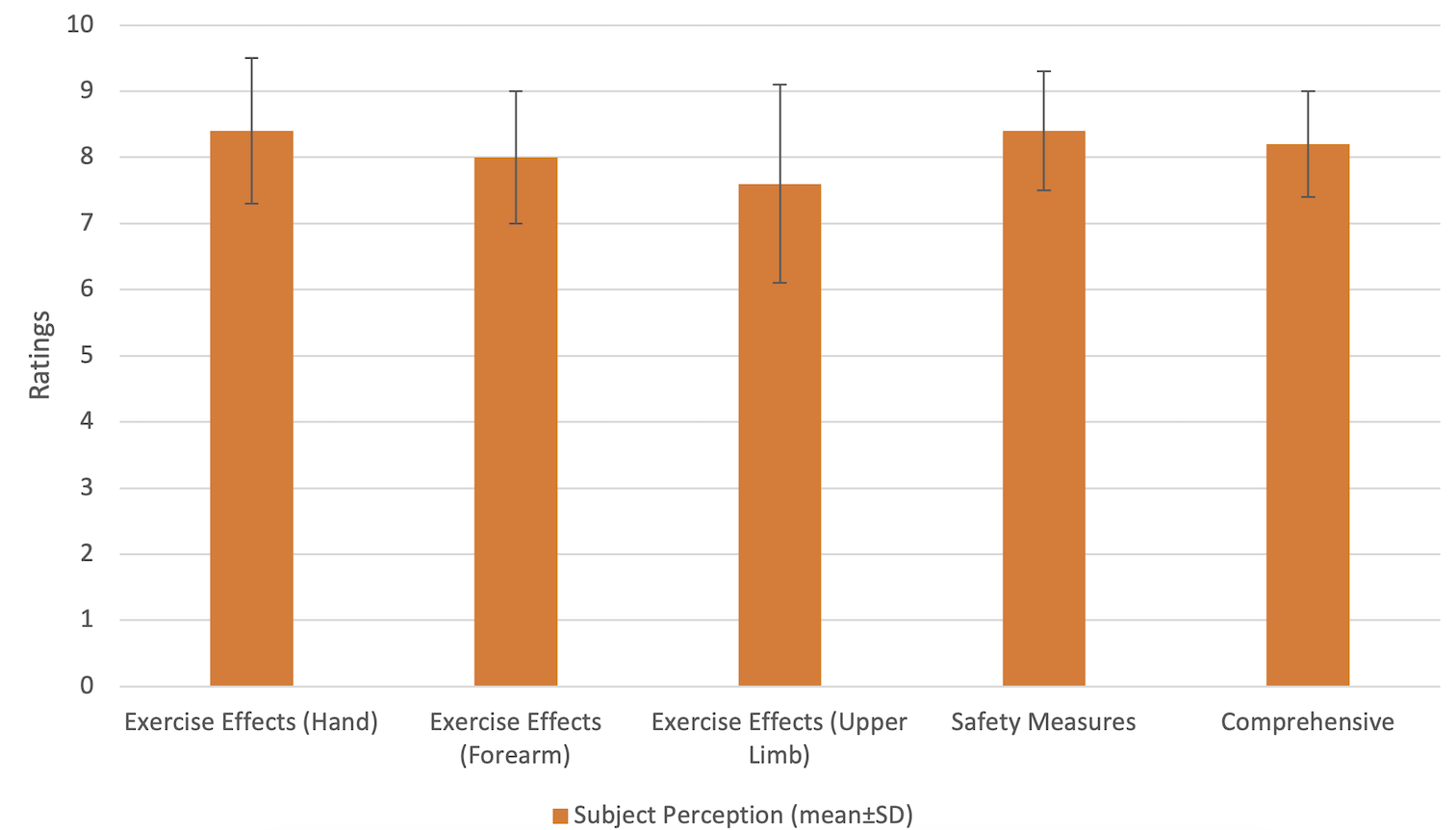}
       \caption{Subject's perception of the collaborative training}
       \label{fig:rating}
   \end{figure}

In addition, we surveyed the subjects on the exercise effects after the experiment. The result was presented in Figure \ref{fig:rating} with a 1-to-10 scale, larger means better. We could conclude from the subject's perspective, this further endorsed the training effects on muscle activation by the designed task. Subjects rated high on the safety, the arm and hand training effect after the task. 
All of them recommend the training on our designed configuration and tasks if required.

%\begin{table}[htbp]
%\centering
%\resizebox{0.5\textwidth}{!}{%
%\begin{tabular}{|c|c|}
%\hline
 %\textbf{Questions} & \textbf{Score (mean $\pm$ SD)} \\
 %\hline
%Exercise effect on hand after the training & 8.4 $\pm$ 1.1 \\ 
%\hline
%Exercise effect for forearm after the training & 8.0 $\pm$ 1.0 \\ 
%\hline
%Exercise effect for upper arm after the training  & 7.6 $\pm $ 1.5 \\ 
%\hline
%Feelings on Safety measures during training & 8.4 $\pm $ 0.9 \\ 
%\hline
%Comprehensive rating & 8.2 $\pm $ 0.8 \\ 
%\hline
%\end{tabular}%
%}
%\caption{Survey Results after completing the collaborative training}
%\label{tab:my-table}
%\end{table}

\section{Conclusion}

We proposed a novel adaptive system for assistive rehabilitation training with Cobot integrating cameras, force sensors, and robot controllers. We examined two representative tasks: a gross level motor training task and a fine level motor training task. In each task, the agent learned a policy from expert demonstrations and can be adjusted according to the subject's physical capabilities. Our work provided an all-in-one platform and demonstrated the feasibility of developing in-home personalized remote upper-limb training system with general-purpose robot for individuals with motor disabilities. 

%This work demonstrates the feasibility of using Cobot for multi-purpose upper-limb rehabilitation training. Immediate future work could be recruiting patients for long-term experiments throughout the subacute stage of rehabilitation training to evaluate the efficacy. Integrating sEMG feedback into the control loop may lead to more effective trajectory planning. More sensors could be used to monitor subject physiological parameters, such as heart rates, electrocardiogram (ECG), and breathing rates, which are valuable for physicians and therapists while conducting a remote training task.

%This work demonstrates the feasibility of training with the subject's preference by using a collaborative robot through imitation learning. Some future work could be recruiting patients for long-term experiments throughout the subacute stage of rehabilitation training to evaluate the efficacy. Integrating sEMG feedback into the control loop may lead to more effective trajectory planning. More sensors could be used to monitor subject physiological parameters, such as heart rates, electrocardiogram (ECG), and breathing rates, which are valuable for physicians and therapists while conducting remote rehabilitation training via telemedicine.

\section*{ACKNOWLEDGMENT}
 This work was supported by the National Key Research and Development Program of China (2021YFE0111800), Beijing Municipal Science and Technology Program (Z211100003521006), STI 2030 - Major Projects (2022ZD0209400), Shuimu Tsinghua Scholar Program (2019SM133). The experiments involving human subjects described in this paper were approved by the IRB of Tsinghua University. The authors would like to thank Dr. Lim Soon Bock and his team for medical advice given.

\end{document}